# Truck drivers and automation: A methodology for identifying and supporting workforce transition in the Australian road freight sector.

A. Bratanova*, C. Mason, D. Evans, E. Schleiger, E. Grimberg, G. Walker, H. Pham and K. Bulled

## Abstract

The transition to autonomous trucks (ATs) is coming, and although it is expected to be gradual, it will create both challenges and opportunities for the existing driver workforce. This paper presents a novel integrated methodology for identifying viable occupational transitions for truck drivers as transport automation advances. Unlike traditional workforce transition analyses that focus primarily on skill similarity, wages, and employment demand, this methodology incorporates four integrated components: task-level automation analysis, skill similarity assessment, labour market conditions analysis, and empirical validation using historical transition patterns.

Applying this methodology to Australian truck drivers shows that while ATs will automate core driving tasks, many non-driving responsibilities will continue requiring human involvement, suggesting occupational evolution rather than wholesale displacement. A skill similarity analysis identifies 17 occupations with high transferability, while labour market analysis reveals significant trade-offs between wage levels and job availability across potential transition pathways.

Key findings indicate that bus and coach driving, along with earthmoving plant operation, emerge as high-priority transition options, offering comparable wages and positive employment growth. Delivery and forklift driving present medium-priority pathways with abundant opportunities but lower wages. A regression analysis of historical transitions confirms that skill similarity, wage differentials, geographic accessibility, and qualification requirements all significantly influence actual transition patterns, with some viable pathways currently underutilised.

The research provides policymakers, industry stakeholders, and educational institutions with evidence-based guidance for supporting workforce adaptation to technological change. The proposed methodology is generalisable beyond trucking to other sectors facing automation, offering a systematic approach for managing inclusive transitions in an era of technological disruption. By moving beyond theoretical skill matching to incorporate real-world labour market dynamics, this methodology enables more targeted and effective workforce planning strategies.

*Corresponding author. Email: Alexandra.bratanova@csiro.au. Postal address: Ecosciences Precinct, 41 Boggo Rd, Dutton Park, Queensland 4102, Australia

**Keywords:** Automation, skills, occupation transition, truck driver, autonomous vehicles, workforce planning, Australia

**JEL classification codes:** J24, J62, O33

**Authors' affiliation:** Commonwealth Scientific and Industrial Research Organisation (CSIRO), Brisbane, Queensland, Australia

**Funding:** CSIRO

**Acknowledgments:** Authors would like to acknowledge reviewers, and colleagues from CSIRO and other organisations who helped check and improve early drafts of this manuscript.

## 1. Introduction

The global transition to transport automation is advancing rapidly, enabled by breakthroughs in artificial intelligence, sensor technology, and computational capabilities. Within this landscape, autonomous trucks (ATs) have emerged as a transformative innovation in the road freight industry, offering the potential to significantly improve efficiency, safety, and environmental sustainability.

The Autonomous Vehicle Industry Association reports that its members collectively achieved over 70 million kilometres of autonomous driving on U.S. roads by 2024 — equivalent to 184 trips to the Moon [1]. In the freight sector specifically, companies such as TuSimple, Plus, Aurora, Volvo, and IVECO have conducted commercial trials around the world demonstrating technological viability and potential economic benefits [2, 3], with reported improvements in fuel efficiency of 10-15% [4] and significant reductions in operational times for long-haul routes [5]. ATs have also begun commercial operations in limited contexts [6].

In Australia, AT trials have progressed more cautiously. The 2022 Transurban trial involved an autonomous truck that drove itself on a limited route over a week. The 2024 trial was originally intended to involve two trucks equipped with automated driving systems operating in autonomous mode during overnight periods with specially trained vehicle supervisors present. However, in response to stakeholder feedback, the trial approach was adapted so that professional drivers maintained control of the vehicles at all times, while the automated driving systems operated in "shadow mode" (a configuration in which the system performs all driving functions in parallel to the human driver but without controlling the vehicle) to collect performance data [7].

This cautious approach reflects the significant economic and safety stakes involved. In Australia, the road freight industry is a vital component of the national economy, contributing AU$164.4 billion (7.9% of GDP) and employing over 204,000 truck drivers [8]. Yet the sector faces persistent structural challenges, including workforce shortages, aging driver population, and mounting pressures to improve safety and reduce environmental impacts [9]. The median age of Australian truck drivers reached 48 years in 2024 [10], notably older than the national average of 40 [11]. Truck driving is labelled as one of Australia's most dangerous jobs [12] — drivers have been found to have a 13-fold greater risk of fatal injuries than workers in other occupations [13]. As of November 2023, more than 3,200 truck driver vacancies were advertised in a single month, signalling ongoing recruitment difficulties [14]. These challenges are not unique to Australia. Globally, over three million truck driver positions remained unfilled in 2023, a figure expected to double by 2028 [15].

While the expected timing of AT adoption remains debated ranging from years to decades [16-21], there is broad agreement that automation will significantly impact the labour market in the road freight sector [21-23]. Proponents argue that ATs are a necessary response to driver shortage [23] and can reduce labour costs [24], while also improving safety by minimising human error [5]. Critics fear that ATs will displace the existing workforce. Given broader barriers to AT adoption (e.g., technological limitations, safety concerns, public trust, cybersecurity risks, regulatory uncertainty, and underdeveloped infrastructure [16, 17]), full displacement is unlikely in the near term; instead, AT deployment is expected to be gradual. It will reshape the occupational landscape and require substantial workforce adaptation. Historical precedents in mining and agriculture demonstrate that, while job losses may be offset by new roles, these roles often require more advanced skills and may be fewer in number. In Australia's mining sector, for example, automation has shifted labour demand from manual tasks to roles in remote operations and systems maintenance [25, 26].

Despite growing recognition of workforce implications of truck automation, existing research has largely focused on the technical feasibility and economic impact, with limited attention to how drivers might transition into new occupations [21]. Where workforce transition studies do exist, they typically rely on three factors: skill similarity, relative wages, and projected employment demand [27, 28]. Such models often neglect real-world factors that affect transition feasibility, such as geographic immobility, educational requirements, and cultural barriers [29]. In the Australian context, research on occupational transitions in road freight has been especially limited. To date, only one study has examined alternative occupations for truck drivers by geographies in the U.S. [21] and none in Australia.

This paper addresses this gap by presenting a methodology for assessing occupational transitions in the road freight sector. Its primary contribution is the development and empirical validation of an integrated set of analytical solutions that extends existing workforce transition analyses in three ways:

- Leverages actual labour force data to reveal observed occupational transitions in practice.
- Synthesises multiple analytical components into a cohesive methodology providing actionable insights.
- Incorporates geographic constraints and historical transition patterns — factors often omitted in standard analyses.

The empirical analysis reveals that Australian truck drivers have skills profiles similar to other occupations likely to be in high demand in the future. Bus and coach driving and earthmoving plant operation emerge as viable alternatives, offering comparable wages and employment growth. Delivery and forklift driving are also in high demand but offer lower median wages, reducing their attractiveness for transitioning truck drivers.

While the empirical analysis focuses on the trucking sector, the methodology is broadly applicable to other industries experiencing technological disruption, providing a systematic approach for supporting inclusive and sustainable workforce transitions.

The remainder of the paper is structured as follows. Section 2 reviews the relevant literature on automation and occupational transitions; Section 3 outlines the methodological approach; Section 4 applies the methodology to the Australian road freight sector; Section 5 presents discussion of key findings, limitations, and implications; and Section 6 concludes.

## 2. Background and literature review

### 2.1 Labour market impacts of automation: Theoretical frameworks

The relationship between technological change and employment has evolved from early concerns about technological unemployment [30] to more nuanced frameworks acknowledging both displacement and reinstatement effects [31]. The task-based approach, pioneered by Autor, Levy, and Murnane [32] and further developed by Acemoglu and Autor [33], provides a valuable framework for understanding automation's differential impacts across occupations by categorising work tasks along dimensions such as routine vs. non-routine and cognitive vs. manual.

In the context of truck driving, this framework suggests that while core driving tasks may be automated, many non-routine tasks like vehicle inspection, cargo management, and customer interaction may still require human involvement [34]. Historical precedent supports this nuanced view: studies by Bessen [35, 36] demonstrate that technological change has often led to industry growth and increased employment, even in directly affected occupations. The example of ATMs and bank tellers illustrates this counterintuitive outcome: despite automation of cash-handling tasks, overall employment of bank tellers increased as branches expanded and roles evolved to focus on customer relationship management [35].

However, automation's distributional effects can be significant. Acemoglu and Restrepo [37] showed that automation can lead to wage polarisation, with middle-skill jobs most affected — raising concerns for truck drivers, who typically earn middle-range incomes. Geographic factors can also lead to concentrated impacts in regions with high employment in automation-susceptible occupations [38].

Regarding the pace of change, recent empirical work on autonomous vehicles and employment presents mixed findings. Gittleman and Monaco [34] conclude that truck driving jobs are likely to persist longer than many predictions suggest, given the complexity of driving tasks and drivers' broader roles beyond vehicle operation. Similarly, Groshen et al. [39] project a gradual transition rather than wholesale displacement. However, Wang et al. [21] present a more nuanced regional analysis, identifying certain contexts where more rapid transitions to ATs could occur, particularly in areas with supportive regulation and infrastructure.

### 2.2 Skills and occupational transitions: Methodological approaches

Understanding the skill composition of occupations is crucial for identifying viable transition pathways. The literature on skill transferability and occupational mobility provides several methodological approaches.

Shaw and Lazear [40] demonstrate that occupation-specific human capital accounts for a significant portion of workers' earnings, highlighting the importance of preserving existing skills in transition design. This concept of "skill distance" between occupations has been operationalised in various ways, with researchers using measures such as Euclidean distance [41] or correlation coefficients [42] to quantify skill similarity. In freight transport research, a study by Van Fossen et al. [27] leveraged Occupational Information Network (O*NET) data to identify potential transition occupations for truck drivers facing displacement due to autonomous vehicles. They highlighted occupations with similar skill requirements, focusing on potential transition destinations that maximise skill transferability. This approach is particularly valuable for identifying potential transition pathways that minimise retraining requirements.

However, successful occupational transitions depend on multiple factors beyond skill similarity. Wage differentials, employment growth, geographic distribution, credentialing requirements, and cultural factors all shape feasibility [43-45]. Human capital theory provides a foundational framework for understanding these complexities. As Becker [46] explains, human capital, comprising the skills, knowledge, and experience accumulated through education and work, is often specific to a given occupation. This specificity can limit transferability and create friction when workers shift to different roles or industries, due to both skill gaps and the sunk costs of role-specific knowledge. These barriers are further compounded by geographic immobility, credentialing requirements, and perceived cultural distance between occupations.

Consequently, even when occupations appear similar in terms of skills, the actual cost and feasibility of transition may vary widely. As Carnevale et al. [47] and Holzer [48] emphasize, effective transition pathway design requires considering both supply-side factors (worker capabilities) and demand-side realities (labour market conditions such as wage differentials, projected job growth, regional availability of opportunities, and actual job openings). This dual perspective ensures that transition recommendations reflect not only theoretical skill compatibility but also practical labour market dynamics. Van Fossen et al. [27] specifically examined truck driver transitions, discovering that while skill profile similarity remains important, drivers also evaluate occupational alternatives based on work activities performed and the industries where occupations are employed, suggesting that successful transition methodologies must capture both quantitative skill matches and qualitative work context factors.

Reflecting these insights, researchers have developed diverse methodologies for identifying suitable workforce transitions, ranging from skill profile similarity measures to assessments of industry viability within regions based on existing workforce capabilities [49-52]. These approaches typically incorporate variables such as automation potential [49, 51], educational requirements [49], projected demand growth [51], and experience requirements [49].

Despite this growing body of work, only one other study has explicitly used real-world occupational transitions data to inform recommendations. Specifically, Howison et al. [52] used machine learning models on employment histories and wage data to identify transitions associated with earnings gains. However, these transitions were assessed narrowly in terms of wage outcomes rather than broader labour market viability.

## 2.3 Managing workforce transitions: Policy and practical approaches

From the policy perspective, managing workforce transitions in the context of technological change requires coordinated action across multiple stakeholders. Surveys of truck drivers in both Australia and the U.S. indicate a preference for jobs with similar skill sets, suggesting that targeted transition pathways could ease this shift [27]. The literature identifies several domains relevant to facilitating transitions:

1. Education and training systems: Targeted, stackable training programs that build on existing capabilities have proven most effective for workers facing displacement [38]. For truck drivers specifically, Yankelevich et al. [53] highlight the potential for training programs focused on developing skills for roles in fleet management, logistics coordination, or remote vehicle operation.

2. Labour market intermediaries: Workforce development organisations, unions, and industry associations play a crucial role in coordinating and supporting transitions [54] In the context of ATs, organisations like the Australian Trucking Association have advocated for proactive workforce planning and transition support [55].

3. Public policy: Regulatory frameworks and social protection systems significantly shape both the pace of technological adoption and the inclusivity of workforce transitions. Regulatory frameworks for autonomous vehicles not only affect safety and infrastructure requirements but also influence the timeline for workforce impacts. Additionally, social protection systems, active labour market policies, and regional development initiatives can mitigate potential negative effects of technological displacement [56, 57].

4. Employer practices: Employers are central to transition management, as companies can directly reskill and redeploy workers for new roles. The International Transport Forum and OECD [58] document examples of transportation companies implementing gradual transitions that incorporate existing drivers into new technological systems, either as operators with enhanced capabilities or in entirely new roles.

Together, the reviewed strands of research underscore that effective transition planning must integrate worker capabilities, labour market realities, and institutional supports. The methodology presented in the next section addresses these needs by combining skill-based analysis with labour market conditions and real-world evidence of occupational mobility.

## 3. Methodology and data

### 3.1 Methodology

The methodology presented in this paper consists of four interconnected analyses designed to identify and validate feasible occupational transitions for truck drivers whose jobs and tasks are likely to be impacted by automation. Together, these are synthesised into a methodology for identifying and supporting feasible transition pathways. Figure 1 presents a visual summary of the methodology.

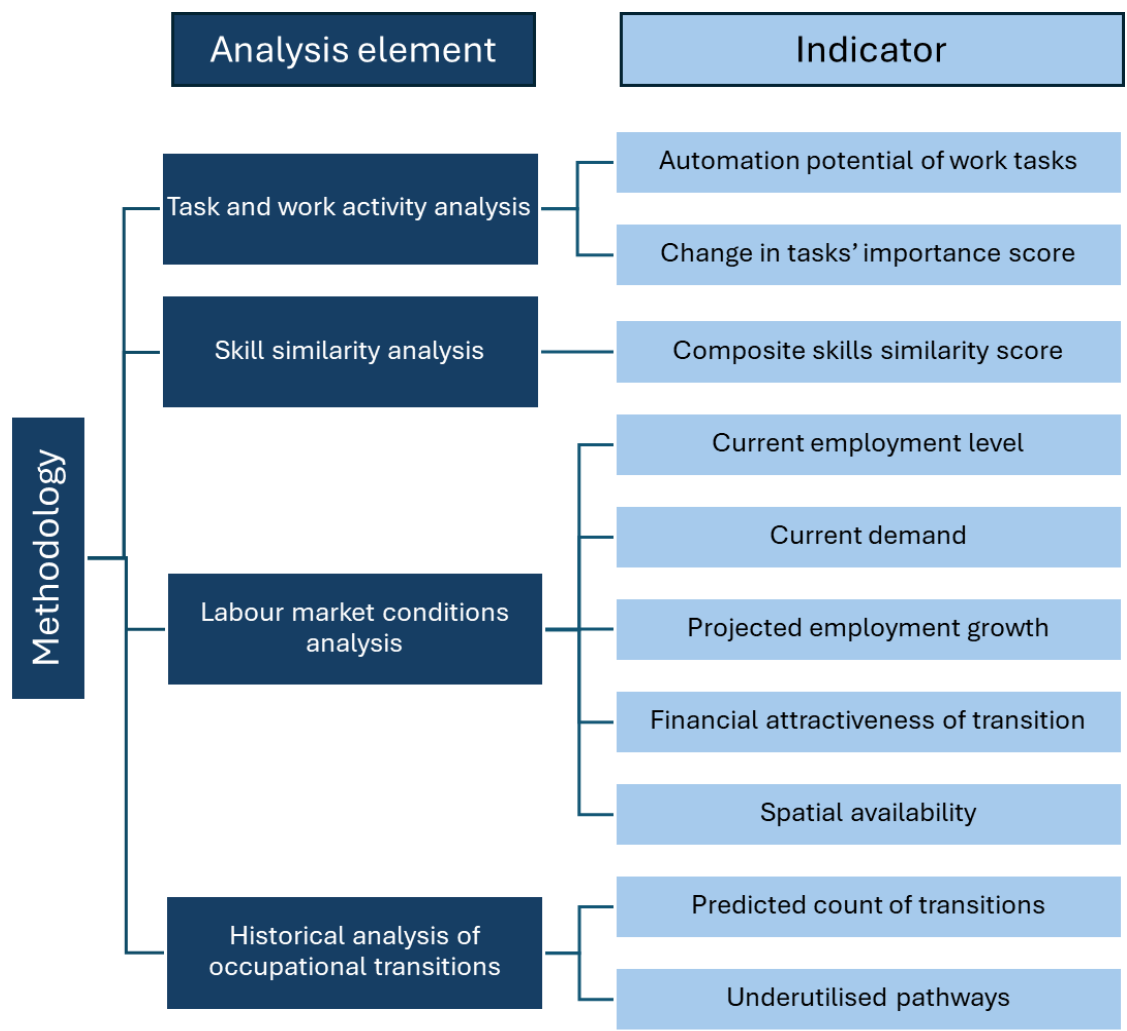

Figure 1. Methodology for identifying and supporting workforce transitions.

**3.1.1 Task and work activity analysis**

This analysis identifies which aspects of the truck-driving occupation are most susceptible to automation and which are likely to remain human-dependent.

The analysis draws on O*NET, a widely used database for examining skill requirements across occupations [59, 60]. Using O*NET data on skills, abilities, knowledge, and work activities, tasks are categorised into three groups:

- Driving tasks likely to be automated by ATs,
- Non-driving tasks that can potentially be automated by other technologies, and
- Tasks likely to continue requiring human involvement.

This granular task-level analysis provides a foundation for understanding the degree of automation exposure and identifies specific skills and capabilities that may remain valuable in a more automated environment. Additionally, the general work activities of truck drivers are examined to understand how their importance may evolve with ATs deployment, highlighting potential areas for skill development and role evolution.

### 3.1.2 Skill similarity analysis

Building on the task analysis, similarity scores are computed for each alternative occupation. This process involves extracting comprehensive skill, ability, and knowledge profiles for truck drivers from O*NET data, comparing these profiles to those of all other occupations, and calculating a composite similarity score that reflects the overall match between truck driving and alternative occupations. The similarity scores range from 0 to 100, where 0 represents the occupations most dissimilar to truck driving and 100 represents a perfect match with the truck-driving occupation. Occupations with similarity scores of 70 or higher are examined. This threshold reflects substantial overlap in skill requirements and relatively low retraining needs. Through this approach, specific occupations are identified, where truck drivers' existing capabilities are most transferable.

### 3.1.3 Labour market conditions analysis

Labour market viability is assessed for each occupation identified through the skill similarity analysis. This assessment includes analyses of:

- Current employment levels, which indicate the overall size of the occupation,
- Recent job postings, which serve as a proxy for current demand,
- Projected employment growth, which indicates future opportunities,
- Median income, which affects the financial attractiveness of transitions, and
- Geographic distribution of employment, which determines the spatial availability of opportunities.

By considering these practical labour market factors, the resulting recommendations are grounded in real-world opportunities rather than theoretical skill matches.

### 3.1.4 Historical analysis of occupational transitions

To validate the identification of feasible transitions, a regression model is developed to predict the annual transition counts from truck driving to other occupations. This model incorporates a set of predictors including skill similarity between occupations, wage differentials, employment levels and growth rates, formal skill level requirements, and geographic distribution of employment. This approach uncovers real-world factors that influence transition patterns beyond skill similarity alone, thereby providing empirical validation of the methodology.

### 3.1.5 Synthesis and recommendation methodology

The four components are then integrated into a decision-support methodology for identifying and prioritising occupational transition pathways. The methodology considers:

- Skill transferability, as measured by skill similarity scores,
- Labour market viability, based on employment levels, growth projections, and job postings,
- Financial feasibility, based on median income comparisons, and
- Geographic accessibility, based on the spatial distribution of employment opportunities.

By integrating these dimensions, an assessment of transition pathways is created that accounts for both worker capabilities and labour market realities.

## 3.2 Datasets and measures

The analysis draws on four data sources to develop a comprehensive understanding of transition feasibility (see Figure 2).

1. O*NET database is the primary data source for the analysis. It provides detailed occupational information, defining 35 skills, 52 abilities, and 33 knowledge types with importance and level scores for each occupation. The database also outlines 41 general work activities rated on a 1-5 importance scale based on worker surveys. While O*NET is U.S.-based, established correspondences between the U.S. Standard Occupation Classification (SOC) and the Australian and New Zealand Standard Classification of Occupations (ANZSCO) enable application to Australian drivers.

2. Australian Bureau of Statistics (ABS) data is incorporated to understand the status and trends in the Australian labour market. The 2021 Census data allows to identify the size and demographic profile of the truck-driving workforce. Labour Force Survey data provides recent estimates of employment by occupation and industry. Income data from the Survey of Income and Housing allows to compare median incomes across occupations. In addition, wage data from the ABS Employee Earnings and Hours survey provides detailed information on wage levels by occupation.

**DOL - U.S. Department of Labour**
- O*NET database

**ABS - Australian Bureau of Statistics**
- 2021 Census
- Labour Force Survey
- Survey of Income and Housing
- Employee Earnings and Hours survey

**JSA - Jobs and Skills Australia**
- Skill requirements
- Five-year employment projections
- Occupation profiles

**NSC - National Skills Commission**
- Internet Vacancy Index

Figure 2. Four data sources for empirical analysis.

3. Jobs and Skills Australia (JSA) data provides employment projections, occupation profiles, and skill requirements.

4 National Skills Commission data provides information on online job advertisements by occupation at the national, state, and regional levels through the Internet Vacancy Index. This data serves as an indicator of current labour market demand for specific occupations.

Together, these datasets provide skill profiles, labour market dynamics, and real-world evidence of occupational mobility for Australian truck drivers.

# 4. Analysis

## 4.1 Task and work activity

The task-level analysis reveals that automation will affect some, but not all, components of truck driving. Table 1 presents 16 tasks performed by truck drivers, indicating whether each task is a core or non-core function and its susceptibility to automation.

Table 1. Truck drivers' detailed work tasks and automation potential.

| Task | Task type | Task status |
|---|---|---|
| Drive trucks to weigh stations before and after loading and along routes in compliance with state regulations. | Core | Automated by ATs |
| Drive trucks with capacities greater than 13 tonnes, including tractor-trailer combinations, to transport and deliver products, livestock, or other materials. | Core | Automated by ATs |
| Manoeuvre trucks into loading or unloading positions, following signals from loading crew and checking that vehicle and loading equipment are properly positioned. | Core | Automated by ATs |
| Drive electric or hybrid-electric powered trucks or alternative fuel-powered trucks to transport and deliver products, livestock, or other materials. | Non-core | Automated by ATs |
| Check all load-related documentation for completeness and accuracy. | Core | Automatable by other technologies |
| Check conditions of trailers after contents have been unloaded to ensure that there has been no damage. | Core | Automatable by other technologies |
| Plan or adjust routes based on changing conditions, using computer equipment, global positioning systems (GPS) equipment, or other navigation devices, to minimize fuel consumption and carbon emissions. | Core | Automatable by other technologies |
| Read bills of lading to determine assignment details. | Core | Automatable by other technologies |
| Check vehicles to ensure that mechanical, safety, and emergency equipment is in good working order. | Core | Human required |
| Obtain receipts or signatures for delivered goods and collect payment for services when required. | Core | Human required |
| Perform basic vehicle maintenance tasks, such as adding oil, fuel, or radiator fluid, performing minor repairs, or washing trucks. | Core | Human required |
| Remove debris from loaded trailers. | Core | Human required |
| Follow appropriate safety procedures for transporting dangerous goods. | Non-core | Human required |
| Follow special cargo-related procedures, such as checking refrigeration systems for frozen foods or providing food or water for livestock. | Non-core | Human required |
| Give directions to labourers who are packing goods and moving them onto trailers. | Non-core | Human required |
| Wrap and secure goods using pads, packing paper, containers, or straps. | Non-core | Human required |

Source: O*NET (2024)

Four of the 16 identified tasks are directly automatable by ATs, four are potentially automatable by other technologies, though economic feasibility may vary. Eight tasks remain human-dependent (e.g., vehicle maintenance, cargo management, safety procedures).

Figure 3 extends this analysis by examining the general work activities of truck drivers and how they are likely to evolve with automation. Horizontal axis indicates the current importance score of each activity. Purple activities represent driving-related tasks likely to be automated by ATs. Green activities represent non-driving tasks that will remain important, such as handling objects, inspecting equipment, communicating with coworkers, and maintaining equipment. Blue activities represent new work activities that may become more important for AT operators or support workers, such as planning and scheduling, processing information, and making decisions.

This task-level analysis reveals three critical insights for transition support. First, ATs primarily automate driving-specific activities while leaving many other aspects of the truck driver's role largely unaffected. Second, many current tasks will remain relevant in an automated environment, providing a foundation for role evolution rather than wholesale displacement. Third, automation introduces new activities that can form the basis of evolving roles for truck drivers transitioning into AT operator or support positions.

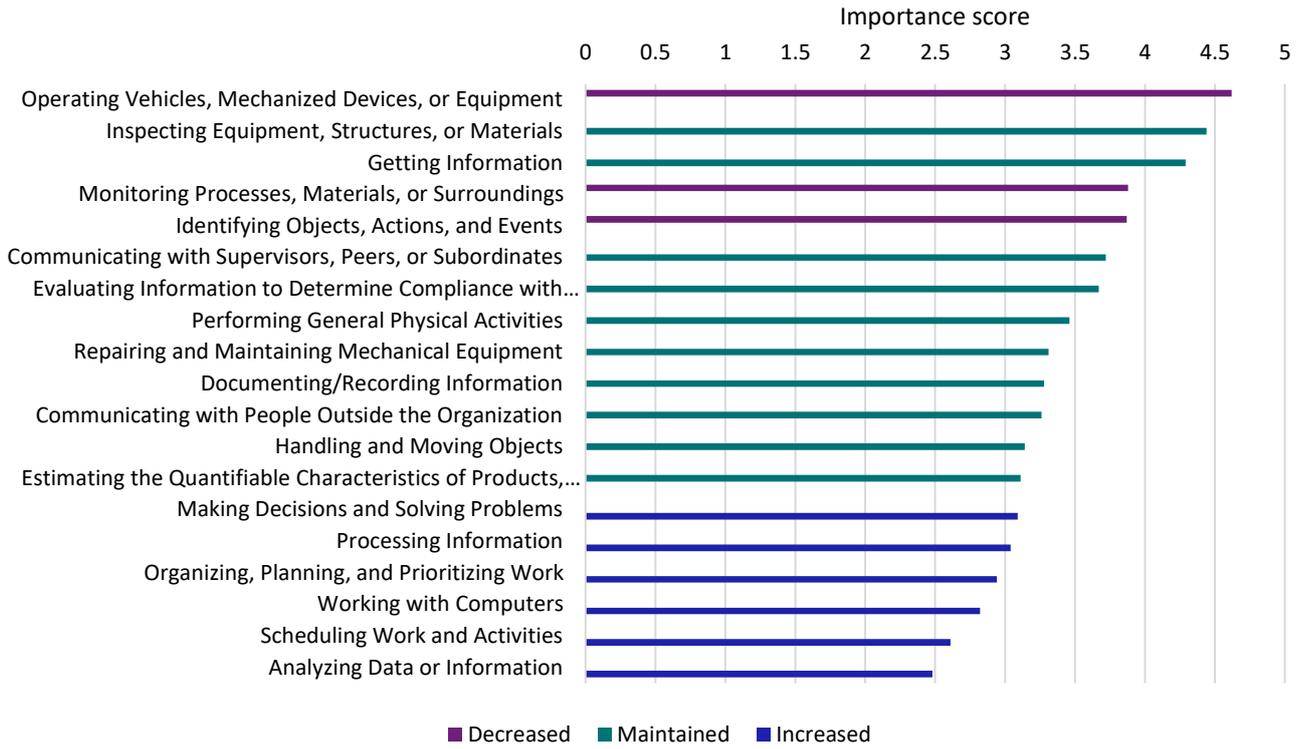

Figure 3. Evolving truck-driver work activities.

## 4.2 Skill similarity

Building on the first analysis, the second stage involves calculating the similarity between truck driving and other occupations (level 4 ANZSCOs) based on the skills, abilities, and knowledge requirements from the O*NET. The difference in capability requirements between truck driving and occupation $j$ is computed as:

$$DIFF_j = \sum_i \left|LEV_{ij} - LEV_{i,truck}\right| + \left|IMP_{ij} - IMP_{i,truck}\right|$$

where $LEV_{ij}$ is the level of the $i$th O*NET capability (skill, ability, or knowledge category) required to work in the $j$th occupation, $LEV_{i,truck}$ is the level of the capability required to work as a truck driver, $IMP_{ij}$ is the importance of the capability for working in the $j$th occupation, and $IMP_{i,truck}$ is the importance of the capability for working as a truck driver. Then, $DIFF_j$ is transformed into a score indicating the similarity in required capabilities between truck driving and occupation $j$:

$$SIM_j = 100\left[1 - \frac{\left(DIFF_j - \min\left(DIFF_j\right)\right)}{\left(\max(DIFF_j) - \min(DIFF_j)\right)}\right]$$

A similarity score of 100 indicates that occupation $j$ has identical capability requirements to truck driving, while a score of 0 indicates the least similarity. A threshold skill similarity score of 70 is used to identify occupations with relatively high capability overlap with truck driving. This threshold is selected as it substantially exceeds the mean and median similarity scores between truck driving and other occupations (38 and 36 respectively) and has been used in prior research to identify feasible occupation transition pathways with relatively low retraining needs [52]. Table 2 presents 17 occupations with similarity scores of 70 or above.

Table 2. Occupations with high skill similarity to truck driving.

| Occupation | Similarity Score |
|---|---|
| Crane, Hoist and Lift Operators | 78 |
| Delivery Drivers | 76 |
| Earthmoving Plant Operators | 75 |
| Bus and Coach Drivers | 75 |
| Train and Tram Drivers | 75 |
| Railway Track Workers | 74 |
| Drillers, Miners and Shot Firers | 74 |
| Other Construction and Mining Labourers | 73 |
| Deck and Fishing Hands | 73 |
| Agricultural, Forestry and Horticultural Plant Operators | 73 |
| Forklift Drivers | 72 |
| Other Mobile Plant Operators | 72 |
| Paving and Surfacing Labourers | 72 |
| Engineering Production Workers | 72 |
| Packers | 71 |
| Metal Engineering Process Workers | 71 |
| Forestry and Logging Workers | 70 |

The scores indicate that truck drivers possess many of the skills, abilities, and knowledge types required in these occupations, suggesting potential for transitions with relatively minimal retraining. Most of these occupations do not have formal educational requirements beyond what truck drivers typically possess, with train and tram drivers being the only exception (requiring a Certificate IV qualification).

Identifying occupations with high skill similarity represents the second stage of the methodology: mapping potential destination occupations that maximise skill transferability. While this offers a starting point for viable transition pathways, skill similarity alone is insufficient. Further refinement requires incorporating practical labour market factors as demonstrated below.

## 4.3 Labour market conditions

To assess transition feasibility, labour market conditions are analysed in four major Australian urban areas: Brisbane, Melbourne, Perth, and Sydney. Table 3 compares the identified occupations by median income, employment levels, and recent employment growth. This multidimensional analysis of labour market conditions reveals several key insights for transition support:

1. Financial viability: Bus and coach drivers and earthmoving plant operators offer median incomes equal to or above the AU$71,500 median income for fulltime truck drivers (aside from bus and coach drivers in Brisbane), making them financially attractive. In contrast, delivery drivers, forklift drivers, and packers offer substantially lower median incomes, potentially limiting their appeal despite abundant opportunities.

2. Employment opportunity: Delivery drivers, bus and coach drivers, and forklift drivers have high levels of employment across all regions, indicating strong current demand. Earthmoving plant operators show moderate demand across the regions.

3.      Future prospects: Bus and coach drivers and earthmoving plant operators show positive annual employment growth between 2016 and 2021, suggesting sustained demand over the medium term. This contrasts with employment declines in some other occupations with high skill similarity over the same period.

4.      Geographic considerations: Significant regional variations in employment opportunities exist, with certain occupations showing stronger demand in specific regions. For example, earthmoving plant operators show particularly high employment levels and strong demand growth in Melbourne, perhaps reflecting recent construction activity in the region.

This labour market analysis forms the third stage of the methodology, refining transition pathways based on practical and region-specific employment considerations. By combining skill similarity with labour market viability, transitions are identified that are not only theoretically feasible but also practically accessible.

**Table 3. Labour market conditions and outcomes for feasible transitions.**

| | Similarity | Income diff. | Employment | Emp. growth |
|---|---|---|---|---|
| **Brisbane** | | | | |
| Crane, Hoist and Lift Operators | 78 | 26,000 | 910 | 1 |
| Delivery Drivers | 76 | -13,000 | 7,039 | 12 |
| Earthmoving Plant Operators | 75 | 0 | 2,147 | 1 |
| Bus and Coach Drivers | 75 | -13,000 | 4,081 | 1 |
| Train and Tram Drivers | 75 | 58,500 | 842 | 6 |
| Railway Track Workers | 74 | 26,000 | 408 | 2 |
| Drillers, Miners and Shot Firers | 74 | 58,500 | 980 | 6 |
| Other Construction and Mining Labourers | 73 | 13,000 | 334 | -1 |
| Deck and Fishing Hands | 73 | 0 | 266 | 1 |
| Agricultural, Forestry and Horticultural Plant Operators | 73 | -13,000 | 276 | -1 |
| Forklift Drivers | 72 | -13,000 | 7,332 | 4 |
| Paving and Surfacing Labourers | 72 | 0 | 498 | 2 |
| Other Mobile Plant Operators | 72 | 0 | 1,039 | -2 |
| Engineering Production Workers | 72 | -13,000 | 1,642 | 1 |
| Packers | 71 | -24,700 | 5,974 | 2 |
| Metal Engineering Process Workers | 71 | -13,000 | 862 | 2 |
| Forestry and Logging Workers | 70 | -13,000 | 72 | 7 |
| **Melbourne** | | | | |
| Crane, Hoist and Lift Operators | 78 | 58,500 | 1,585 | 5 |
| Delivery Drivers | 76 | -13,000 | 13,033 | 15 |
| Earthmoving Plant Operators | 75 | 13,000 | 3,651 | 6 |
| Bus and Coach Drivers | 75 | 0 | 5,297 | 1 |
| Train and Tram Drivers | 75 | 58,500 | 2,967 | 4 |
| Railway Track Workers | 74 | 58,500 | 742 | 13 |
| Drillers, Miners and Shot Firers | 74 | 13,000 | 532 | 6 |
| Other Construction and Mining Labourers | 73 | 26,000 | 464 | 1 |
| Deck and Fishing Hands | 73 | 26,000 | 164 | -4 |
| Agricultural, Forestry and Horticultural Plant Operators | 73 | -13,000 | 360 | 2 |
| Forklift Drivers | 72 | -13,000 | 14,952 | 3 |
| Paving and Surfacing Labourers | 72 | 0 | 854 | 5 |
| Other Mobile Plant Operators | 72 | 13,000 | 1,350 | -4 |
| Engineering Production Workers | 72 | -13,000 | 3,383 | 3 |
| Packers | 71 | -24,700 | 13,343 | 4 |
| Metal Engineering Process Workers | 71 | -13,000 | 1,590 | 0 |
| Forestry and Logging Workers | 70 | -13,000 | 130 | 3 |
| **Perth** | | | | |
| Crane, Hoist and Lift Operators | 78 | 26,000 | 895 | 1 |
| Delivery Drivers | 76 | -13,000 | 5,157 | 9 |
| Earthmoving Plant Operators | 75 | 13,000 | 1,698 | 1 |
| Bus and Coach Drivers | 75 | 0 | 2,938 | 1 |
| Train and Tram Drivers | 75 | 26,000 | 591 | 4 |
| Railway Track Workers | 74 | 26,000 | 327 | 7 |
| Drillers, Miners and Shot Firers | 74 | 58,500 | 4,066 | 15 |
| Other Construction and Mining Labourers | 73 | 13,000 | 969 | 10 |
| Deck and Fishing Hands | 73 | 26,000 | 384 | 0 |
| Agricultural, Forestry and Horticultural Plant Operators | 73 | -13,000 | 157 | -2 |
| Forklift Drivers | 72 | -13,000 | 4,545 | 3 |
| Paving and Surfacing Labourers | 72 | 0 | 293 | 2 |
| Other Mobile Plant Operators | 72 | -13,000 | 706 | -4 |
| Engineering Production Workers | 72 | 0 | 1,953 | 2 |
| Packers | 71 | -24,700 | 3,887 | 3 |
| Metal Engineering Process Workers | 71 | -13,000 | 884 | 7 |
| Forestry and Logging Workers | 70 | -13,000 | 47 | -2 |
| **Sydney** | | | | |
| Crane, Hoist and Lift Operators | 78 | 26,000 | 1,519 | 1 |
| Delivery Drivers | 76 | -13,000 | 13,621 | 12 |
| Earthmoving Plant Operators | 75 | 13,000 | 2,753 | 0 |
| Bus and Coach Drivers | 75 | 0 | 8,112 | 0 |
| Train and Tram Drivers | 75 | 58,500 | 2,171 | 7 |
| Railway Track Workers | 74 | 26,000 | 835 | 6 |
| Drillers, Miners and Shot Firers | 74 | 58,500 | 1,397 | 2 |
| Other Construction and Mining Labourers | 73 | 13,000 | 483 | -6 |
| Deck and Fishing Hands | 73 | 26,000 | 435 | -3 |
| Agricultural, Forestry and Horticultural Plant Operators | 73 | -13,000 | 223 | 0 |
| Forklift Drivers | 72 | -13,000 | 13,948 | 2 |
| Paving and Surfacing Labourers | 72 | 0 | 583 | -3 |
| Other Mobile Plant Operators | 72 | 13,000 | 1,828 | -3 |
| Engineering Production Workers | 72 | -13,000 | 2,679 | 3 |
| Packers | 71 | -24,700 | 12,204 | 1 |
| Metal Engineering Process Workers | 71 | -13,000 | 1,089 | -1 |
| Forestry and Logging Workers | 70 | -13,000 | 34 | 0 |

## 4.4 Historical analysis of occupational transitions

To validate the identification of feasible transition pathways, a regression analysis is conducted to predict the annual count of transitions from truck driving to alternative occupations. The model incorporates multiple hypothesised factors, enabling identification of those most strongly associated with actual occupational mobility.

A negative binomial type-1 (NB1) generalised linear mixed model (GLMM) is used to predict $y_{jt}$ - the number of truck drivers transitioning into destination occupation $j$ in year $t$. This model is given by:

$$y_{jt} \mid X_{jt}, b_j, \mu_t \sim NB1(\mu_{jt}, \alpha)$$

$$Var(y_{jt} \mid \mu_{jt}, b_j, u_t) = \mu_{jt}(1 + \alpha)$$

$$\log(\mu_{jt}) = \beta_0 + \beta_1 SIM_j + \beta_2 INC_{jt} + \beta_3 EMP_{jt} + \beta_4 SPATIAL_{jt} + \beta_5 QUAL_j + b_j + u_t + \log(E_t)$$

$$b_j \sim N(0, \sigma^2_{occ})$$

$$u_t \sim N(0, \sigma^2_{year})$$

Where: $E_t$ is the origin exposure (the number of truck drivers available to transition in year $t$; this enters the model as an offset); $\beta_0, \ldots, \beta_6$ are regression weights to be estimated, $b_j$ and $u_t$ are random intercept terms to capture unobserved heterogeneity across destination occupations and years respectively, and $\log(E_t)$ is an offset term. The definitions of the predictors are provided in Table 4.

Table 4. Definitions of the predictors in the regression model.

| Predictor | Definition |
|---|---|
| $SIM_j$ | The similarity in skill requirements between truck driving and occupation $j$, based on O*NET data. This predictor indicates the suitability of truck drivers for employment in occupation $j$ in terms of meeting the new occupation's skill requirements. |
| $INC_{jt}$ | $INC_{jt} = INC_{jt} - INC_{truck,t}$ is the difference in median full-time incomes between occupation $j$ ($INC_{jt}$) and truck driving ($INC_{truck,t}$) in year $t$, based on Australian census data from 2016 and 2021 (linearly extrapolated between census years). This predictor indicates the average financial incentive for a truck driver to transition into occupation $j$. |
| $EMP_{jt}$ | The number of workers in occupation $j$, which is a proxy for the number of employment opportunities in the $j$th occupation. |
| $SPATIAL_{jt}$ | $SPATIAL_{jt} = -\sum_r |EMP_{r,truck} - EMP_{r,j}|$ is the spatial alignment of employment in truck driving and occupation $j$ across Australian labour markets (level 4 statistical areas), based on Australian census data from 2016 and 2021 (linearly extrapolated between census years). Here, $EMP_{r,truck}$ is the $r$th spatial labour market's share of national employment in truck driving and $EMP_{r,j}$ is its share of national employment in the $j$th occupation. Higher values of $SPATIAL_{jt}$ indicate greater spatial alignment of employment in truck driving and the $j$th occupation, suggesting greater availability of nearby employment opportunities in the $j$th occupation for truck drivers. |
| $QUAL_j$ | Takes the value 1 if the $j$th occupation's formal skill level requirement is equal to or less than truck driving (increasing the feasibility of truck drivers transitioning into the occupation) and 0 otherwise. |

This model is fitted to data on all annual transitions from truck driving to other occupations that occurred in Australia between financial years 2016-17 and 2021-22 (see Table 5 for descriptive statistics; sample size $n = 2,052$). The NB1-GLMM was selected as it produced the lowest AIC amongst the candidate models fitted, providing the best balance between fit and parsimony. The alternative models fitted were: Poisson and quasi-Poisson GLMs, which showed strong overdispersion; a NB1 generalised linear model (GLM), where the variance increases linearly with the mean; a NB2-GLM, where the variance increases quadratically with the mean; a NB2-GLMM; and a NB1-GLMM with no random intercept term for year. Each estimated coefficient $\hat{\beta}_1, \ldots, \hat{\beta}_6$ had the same sign $(+/-)$ under all the fitted models, indicating that the reported estimates are robust to model specification.

Table 5. Descriptive statistics for the response variable and predictors.

| Variable | Mean | SD | Min | Max |
|---|---|---|---|---|
| Transitions count ($y_{jt}$) | 17.2 | 52.8 | 0 | 691 |
| $SIM$ | 18.9 | 6.0 | 6.5 | 30.5 |
| $INC$ ($) | 11,569 | 32,051 | -26,910 | 128,932 |
| $EMP$ | 28,455 | 48,389 | 241 | 378,127 |
| $SPATIAL$ (standardised) | 0.0 | 1.0 | -2.9 | 2.0 |
| $QUAL$ | 0.3 | 0.5 | 0 | 1 |

Table 6 provides the regression estimates for the fixed effect terms in the model. The last column shows the estimated incidence rate ratios (IRRs). Each IRR is the multiplicative factor by which the transition rate from truck driving to another occupation increases for a one-unit increase in the predictor, holding the other predictors fixed. Table 6 reports changes in IRR associated with a two-SD increase in the continuous predictors to make the estimates comparable to the binary predictor $QUAL_j$ switching from zero to one, which represents to a 2.16 SD increase (i.e., all predictors aside from $QUAL_j$ were divided by two-SDs prior to fitting the model). That is, the displayed estimates correspond to relatively large increases in the predictors to put them on a comparable scale to the binary variable. The estimates indicate that truck drivers tended to transition into occupations with high skill similarity, higher median income, many job opportunities, geographically accessible job opportunities, and the same or lower formal skill level requirements, validating the focus on these factors in identifying feasible transition pathways.

Table 6. Regression estimates for the NB1-GLMM fixed effect terms.

| Term | Estimate* | Std. Error | p-value | IRR* (95% CI) |
|---|---|---|---|---|
| $SIM$ | 1.571 | 0.243 | <0.001 | 4.81 (2.99-7.74) |
| $INC$ | 0.570 | 0.133 | <0.001 | 1.77 (1.36-2.30) |
| $EMP$ | 0.908 | 0.099 | <0.001 | 2.48 (2.04-3.01) |
| $SPATIAL$ | 1.828 | 0.183 | <0.001 | 6.22 (4.34-8.91) |
| $QUAL$ | 0.694 | 0.243 | 0.004 | 2.00 (1.24-3.22) |
| Constant | -11.687 | 0.134 | <0.001 | |

*Note: * Estimates and IRRs shown for a two-SD increase*

In the estimated model, the occupation intercepts have variance 2.479 (SD 1.575) on the log scale, indicating substantial between-occupation heterogeneity in the number of transitions. The between-year heterogeneity is much smaller, with the year intercepts having an estimated variance of 0.008 (SD 0.091). The estimated dispersion parameter is $\alpha = 1.33$, indicating moderate extra-Poisson variability in counts after accounting for the covariates and random effects.

### 4.5 Integrating the methodology for workforce transition

The synthesis of the four analyses (task evolution, skill similarity, labour market viability, and historical patterns) creates an integrated methodology for identifying and supporting feasible occupational transitions for truck drivers facing automation. Table 7 presents this integrated assessment for the top six occupations.

Table 7. Integrated analysis for the top six transition occupations.

| Occupation | Skill similarity score | Percentage of truck driver median income (2021 AU$) | Recent growth | Geographic accessibility | Formal qualification gap | Recommended priority |
|---|---|---|---|---|---|---|
| Bus/coach driver | 75 | 100% (Melbourne, Perth, Sydney) <br> 82% (Brisbane) | Positive | High | Minimal | High |
| Earthmoving plant operator | 75 | 118% (Melbourne, Perth, Sydney) <br> 100% (Brisbane) | Positive | Moderate | Minimal | High |
| Delivery driver | 76 | 82% (all locations) | Positive | Very High | None | Medium-High |
| Forklift driver | 72 | 82% (all locations) | Positive | High | None | Medium-High |
| Train/tram driver | 75 | 180% (Brisbane, Melbourne, Sydney) <br> 136% (Perth) | Positive | Low | Substantial | Medium-Low |
| Crane operator | 78 | 136% (Brisbane, Perth and Sydney) <br> 180% (Melbourne) | Positive | Low | Moderate | Medium-Low |

This integrated assessment prioritises transition pathways by considering multiple feasibility dimensions rather than a single factor and identifies three tiers of recommended pathways.

High-priority pathways include:

1. Bus and coach drivers: With high skill similarity (75), comparable median wages (100% of the truck driver median in Melbourne, Perth and Sydney and 82% in Brisbane), strong job availability across all regions, and minimal qualification barriers, this represents the most immediately viable transition for truck drivers seeking alternatives due to automation. The alignment of tasks and work contexts also facilitates this transition.

2. Earthmoving plant operators: Despite moderate current demand, this occupation offers wages slightly higher than truck driving (118% of the truck driver median in Melbourne, Perth and Sydney and 100% in Brisbane), positive recent employment growth, and high skill transferability (similarity score of 75).

Medium-priority pathways include:

3. Delivery drivers: Although offering lower wages (82% of the truck driver median in all locations), the very high level of job opportunities, strong skill similarity (76), and absence of formal qualification requirements make this a viable transition, particularly for truck drivers prioritising employment security over maximum income.

4. Forklift drivers: Similar to delivery drivers, forklift driving offers high job availability but lower wages (82% of the truck driver median in all locations). The high skill similarity (72) and minimal retraining requirements make this a practical transition despite the wage differential.

Lower-priority pathways include:

5. Train and tram drivers: Despite offering higher wages (180% of the truck driver median in Brisbane, Melbourne and Sydney and 136% in Perth) and high skill similarity (75), the limited job availability, geographic concentration of opportunities, and substantial qualification requirements (Certificate IV) create significant barriers to this transition.

6. Crane operators: The limited availability of job opportunities would constrain the number of transitions into this occupation, despite offering higher wages (136% of the truck driver median in Brisbane, Perth and Sydney and 180% in Melbourne) and high skill similarity (78).

There is also a group of underutilised pathways, such as general clerks. Qualitative analysis through surveys or interviews would be needed to understand why these relatively well-paid, high-demand roles are not being filled by truck drivers. This would help identify the underlying factors behind the low transition rate and assess whether they could be addressed through interventions such as targeted training or promotion.

This integrated assessment demonstrates how our methodology moves beyond skill-similar occupations to provide a nuanced understanding of transition feasibility across multiple dimensions of practical viability. The resulting prioritisation supports more targeted and effective strategies for assisting truck drivers as they navigate the evolving landscape of transport automation.

## 5. Discussion and implications

### 5.1 Main findings and contribution

The analysis provides key insights into how ATs may reshape workforce dynamics and inform transition planning. It shows that automation is expected to transform driving tasks while leaving many non-driving responsibilities (such as vehicle inspection, cargo management, and customer interaction) largely intact, indicating role evolution rather than displacement. The analysis also reveals that truck drivers possess transferable skills relevant to multiple occupations requiring minimal retraining. Labour market analysis shows trade-offs between wage levels and job availability, underscoring the need for tailored transition strategies that reflect regional demand and income priorities [21]. Historical analysis confirms that actual transitions are shaped by skill similarity, wage differentials, qualifications, and geographic accessibility, revealing practical barriers that targeted interventions could help overcome.

The approach developed here contributes to the study of occupational transitions under technological change by offering several advantages over traditional methods as outlined in Table 8.

Table 8. Methodological advantages of presented approach.

| Feature of the method | Description |
|---|---|
| Multidimensional assessment | By considering factors beyond skill similarity, including wage comparisons, job availability, geographic distribution, and historical patterns, this approach provides a more realistic assessment of transition feasibility. This multidimensional analysis helps identify pathways that are not only theoretically possible but practically accessible given real-world constraints. |
| Empirical validation | The incorporation of historical transition data provides a reality check on the initial theoretical predictions, revealing which transitions occur in practice and highlighting potential barriers or facilitators not captured by skill mapping alone. The empirical validation ensures that the recommendations reflect the complex factors that influence workforce mobility. |
| Geographic sensitivity | By explicitly considering the spatial distribution of employment opportunities, our approach acknowledges the geographic constraints that often limit occupational transitions. This spatial dimension is frequently overlooked in standard transition analyses but represents a critical factor in determining which pathways are truly viable for workers in specific regions. |
| Actionable prioritisation | The integration of multiple factors into a cohesive assessment framework enables the prioritisation of transition pathways based on overall feasibility rather than any single dimension. This prioritisation provides clearer guidance for stakeholders developing interventions to support workforce transitions. |

The methodology is also generalisable. Beyond truck drivers, it can support workforce planning in sectors facing automation, digitalisation, or other impacts of transformative technologies. Some examples include coal power workers moving to renewable energy, manufacturing workers affected by robotics, retail staff displaced by e-commerce, or healthcare workers adapting to diagnostic AI.

The adaptability of developed integrated methodology stems from its modular structure: distinct analytical components can be tailored to specific contexts, while the integrated supply–demand approach is preserved. This makes the methodology a versatile tool for developing realistic, context-specific workforce solutions across diverse industries.

## 5.2 Policy and practical implications

The findings have important implications for supporting workforce transitions in the road freight sector. For policymakers, the results point to the value of targeted programs that build on existing skills, while also improving the visibility of viable alternatives, addressing regional disparities, and ensuring that workforce considerations are integrated into the regulation of ATs.

Industry and employers can draw on the methodology for proactive workforce planning by aligning automation rollouts with natural attrition, creating internal redeployment pathways, and addressing skill gaps through targeted training and partnerships with related sectors. Educational institutions can play a critical role here by prioritising modular, stackable training programs aligned with employer needs, supported by recognition of prior learning and flexible delivery models that extend access to drivers in regional areas.

For truck drivers, the methodology offers a tool to support career development, highlighting skills that complement automation and guiding the pursuit of additional certifications in high-priority fields such as bus driving or earthmoving. A detailed summary of stakeholder implications is provided in Appendix.

## 5.3 Limitations and future research directions

While the analysis applied in this paper provides valuable insights into potential transition pathways for truck drivers, several limitations should be acknowledged:

- The O*NET database, representing U.S. occupational definitions, may not perfectly align with Australian contexts. Future research could use or develop regionally specific skill taxonomies to enhance precision of similarity measures. Australia's current occupational skills taxonomy, the Australian Skills Classification (ASC), contains only ten competencies (e.g., numeracy, teamwork, writing etc.), making it too high-level to accurately measure skills similarity between occupations [61]. Nevertheless, the skill similarity scores between truck driving and other occupations under O*NET are highly correlated with the corresponding competency similarity scores under the ASC, with a Pearson correlation coefficient of 0.86.

- The focus on occupation-level characteristics does not account for intra-occupational differences in skills, education, demographics (e.g., age), preferences, and circumstances across truck drivers. Future research could incorporate a more personalised approach considering career stage, educational background, and personal priorities.

- The analysis doesn't consider factors such as employer preferences and hiring criteria. Future research could incorporate employer surveys or job posting analyses to understand hiring requirements and entry barriers for better representation of demand-side drivers.

- The timeline for widespread AT adoption remains uncertain, affecting transition urgency and scale. Future research could develop scenario-based approaches accounting for different adoption trajectories. Additionally, more detailed geographic analysis could provide insights into local transition opportunities, particularly in rural areas where freight transport is a significant employer.

- More detailed sub-occupation skill mapping could identify specialisations within truck driving that are differentially exposed to automation, enabling more targeted assistance for drivers with varied experience profiles. Integrating this approach with broader economic impact models would support a comprehensive understanding of how AT adoption affects regional economies and labour markets, including indirect effects and new job creation. Examination of how regulatory, educational, and labour market systems influence transition outcomes would further clarify institutional factors that enable effective workforce adaptation.

These limitations point to several avenues for future research that could extend and refine the presented methodology. By addressing these limitations and pursuing these research directions, understanding of how to support effective workforce transitions under technological disruption can be substantially enhanced.

# 6. Conclusion

The methodology presented in this paper advances workforce transition research by providing an integrated approach to analysing occupational change under technological disruption. Developed in the context of Australian truck drivers, it extends standard analyses by integrating skill similarity with labour market dynamics, geographic considerations, and empirical validation. In doing so, it contributes to the broader field of labour and automation studies by offering a transferable, evidence-based methodology for assessing workforce transitions across sectors affected by emerging technologies.

The adoption of ATs represents a pivotal transformation in road freight with significant workforce implications. The findings indicate that this transition will occur gradually, shaped by technological, regulatory, and societal factors. While core driving tasks are expected to be automated, many non-driving responsibilities, such as vehicle inspection, cargo management, and customer interaction, will remain human-dependent, signalling occupational evolution rather than displacement. Several viable transition pathways were identified: bus and coach driving and earthmoving plant operation demonstrate strong skill transferability and comparable wages, while delivery and forklift driving offer more immediate employment opportunities at lower income levels.

These findings have important policy and practical implications. Effective workforce transition will rely on coordinated action among policymakers, industry, education providers, and workers. Transition strategies should build on existing skills while addressing regional and labour market constraints to ensure inclusive and sustainable adaptation. By aligning technological progress with human capabilities, potential displacement can be transformed into opportunity, ensuring that automation enhances productivity while supporting equitable workforce outcomes.

**Declaration of Generative AI and AI-assisted technologies in the writing process:**

During the preparation of this work the authors used ChatGPT4 and Claude in order to improve readability and language of this manuscript. After using these tools, the authors carefully reviewed and edited the content as needed and take full responsibility for the content of the publication.

# Appendix: Detailed Implications for stakeholders

| Stakeholder group | Key implications | Illustrative actions |
|---|---|---|
| Policymakers | Transition programs should build on drivers' existing skills; importance of visibility of alternatives; need for regional tailoring; regulation should integrate workforce transition. | Develop targeted passenger transport certification programs (bus/coach). Provide short courses in earthmoving equipment operation. Create occupation transition maps with certification, wage, and regional availability information. Implement regional strategies in areas with high truck driver concentrations but limited demand. Require AT-deploying companies to contribute to retraining initiatives. |
| Industry & employers | Use findings for proactive workforce planning; demographic profile allows gradual rollout; leverage task evolution for internal redeployment; support sectoral collaboration | Align automation rollout with natural attrition. Transition drivers internally into fleet management, logistics coordination, or remote vehicle operation. Redeploy drivers into bus/coach roles (92 similarity score). Provide targeted training in digital literacy, data interpretation, and systems management. |
| Educational institutions | Training should be modular, stackable, and accessible; recognise prior learning; align programs with employer needs | Develop credential programs building on existing skills. Offer intensive bridging programs between truck and bus driving qualifications. Partner with employers (e.g., driving schools and public transport operators) for hands-on experience. |
| Truck drivers | Use insights to proactively plan career development; focus on enduring skills; evaluate transition options systematically; highlight transferable skills | Develop skills that complement automation (vehicle maintenance, cargo management, customer service, technological literacy). Pursue certifications in high-priority fields (bus driving, earthmoving) for younger drivers. Emphasise transferable experience (safety awareness, route optimisation, vehicle maintenance, customer service) when applying to destination roles |